\documentclass[10pt]{article}

\usepackage{url} 
\usepackage{graphicx}
\usepackage{amsfonts}
\usepackage{hyperref}
\usepackage{color}
\usepackage{amsthm,comment}
\usepackage{amsmath}

\theoremstyle{plain}

\theoremstyle{definition}
\newtheorem{defn}{Definition}[section]

\theoremstyle{remark}

{\baselineskip20pt\sloppy}

\begin{document}

\title{\vspace{-1cm}Evaluating Network Inference Methods in Terms of Their Ability to Preserve the Topology and Complexity of Genetic Networks\thanks{Dr. Ali Masoudi-Nejad, sole Guest Editor for this paper.\newline
Corresponding authors: \{Narsis.Kiani,Hector.Zenil\}@ki.se}}

\author{Narsis A. Kiani$^*$, Hector Zenil$^*$, Jakub Olczak and Jesper Tegn\'er\\
Unit of Computational Medicine, Science For Life Laboratory\\(SciLifeLab), Center for Molecular Medicine, Department\\of Medicine Solna, Karolinska Institute, Stockholm, Sweden.
}

\date{}

\maketitle

\begin{abstract}
Network inference is a rapidly advancing field, with new methods being proposed on a regular basis. Understanding the advantages and limitations of different network inference methods is key to their effective application in different circumstances. The common structural properties shared by diverse networks naturally pose a challenge when it comes to devising accurate inference methods, but surprisingly, there is a paucity of comparison and evaluation methods. Historically, every new methodology has only been tested against \textit{gold standard} (true values) purpose-designed synthetic and real-world (validated) biological networks. In this paper we aim to assess the impact of taking into consideration aspects of topological and information content in the evaluation of the final accuracy of an inference procedure. Specifically, we will compare the best inference methods, in both graph-theoretic and information-theoretic terms, for preserving topological properties and the original information content of synthetic and biological networks. New methods for performance comparison are introduced by borrowing ideas from gene set enrichment analysis and by applying concepts from algorithmic complexity. Experimental results show that no individual algorithm outperforms all others in all cases, and that the challenging and non-trivial nature of network inference is evident in the struggle of some of the algorithms to turn in a performance that is superior to random guesswork. Therefore special care should be taken to suit the method to the purpose at hand. Finally, we show that evaluations from data generated using different underlying topologies have different signatures that can be used to better choose a network reconstruction method.\\

\end{abstract}

\section{Introduction}

Many real-world networks, such as complex technological and social networks, belong in the category of so called 'complex networks', and display a number of the properties that govern the formation and evolution of complex networks~\cite{bar1,bar2,bar3}. Studying biological networks encompasses network analysis, comparisons, modeling, and alignments aimed at discovering a relationship between network topology on the one hand and biological function and disease on the other. The accurate inference of networks from biological data is an open challenge, and over the past few years it has developed into a broad field of study, driven by the application of ever more sophisticated techniques.

The \textit{Dialogue for Reverse Engineering Assessment and Methods} (DREAM) challenge~\cite{sto1,sto2} has fostered significant progress. The DREAM challenge aims to fairly compare the strengths and weaknesses of inference methods. Network inference methods have complementary pros and cons under different conditions. Ideally, the validation and interpretation of GRN models must keep pace with new knowledge and experimental data available for modeling, and thus it is important to illustrate all aspects and capacities of a network inference method.

Comparing large cellular networks, which is analogous to genetic sequence comparison, will revolutionize biological understanding. However, comparing large networks is computationally infeasible due to the NP-completeness of the underlying subgraph isomorphism problem. Thus, large network analyses and comparisons rely on heuristics, commonly called network parameters or properties. Furthermore our understanding of gene regulatory networks is still only partial~\cite{altay,fgen}.

Generally, researchers produce artificial networks and simulated data for method assessment. Synthetic data do not usually reflect the complexity of a real biological system if no prior biological information is introduced. Although the exact details may differ, most methods of evaluation of network reconstruction consider the sensitivity, specificity, precision, and in some cases, a receiver operating characteristic (ROC) curve in illustrating the performance of a method. The analysis of biological networks has led to the realization that the architecture of these networks shares many features with other complex networks. They show non-trivial topological properties such as modular structure and long-tail degree distribution~\cite{new}.

The common structural properties shared by diverse networks naturally pose a challenge when it comes to devising more accurate inference methods capable of preserving them. Surprisingly, there has been no evaluation or comparison of different models from this point of view.

Understanding the advantages and limitations of different network inference methods is necessary for their effective application in specific circumstances. In this paper we address this question, evaluating the similarity between the structural features of a true network and those of an inferred network. We have chosen 6 different inference algorithms from among the best-performing algorithms in past DREAM challenges. These methods have been studied using statistical performance measures such as the F-score~\cite{fgen} or area under the receiver operator curve (AUROC)~\cite{nar1,Hus}. Attempts have been made to consider aspects of the overall properties of an inferred network other than the specific number of false and true positive/negative edge inference cases~\cite{siegenthaler}.

In this article we analyze network inference methods, employing topological measures and indices in combination with ensemble data in order to assess their performance. An effective similarity metric is needed for scoring network inference methods, one which, given two complex networks, evaluates the degree of similarity between their structural features, beyond just looking at individual numbers. We borrow ideas from gene set enrichment analysis (GSEA)~\cite{GSEA2005, bet, nar2} to formulate an intelligent method which we offer as a new way to measure the topological similarity of two complex networks. We benchmark them using synthetic transcriptional networks proposed by Mendes et al.~\cite{Mendes2003} using a Hill equation method. These networks consist of 100 ``genes'' with 200 connections among them, exemplifying three different topologies: Erd\"{o}s-R\'enyi~\cite{erdos}, small world (SW)~\cite{WattsStrogatz1998} and scale-free~\cite{new} (SF). Mendes et al. have used these networks with well-defined topologies to run \textit{in-silico} experiments simulating real laboratory gene expression values. We compared ARACNe~\cite{ARACNe2004}, CLR (Context Likelihood of Relatedness~\cite{CLR2007}, GENIE3~\cite{GENIE2010}, INFERELATOR~\cite{Inferelator2010}, TIGRESS~\cite{TIGRESS2012} and Correlation on the basis of diameter, average shortest path length, clustering and centrality scores.

\section{Methods}
Six different network inference algorithms are considered in this study and will be discussed in the following section. Table~\ref{T1} summarizes the differences between the models analyzed.

\subsection{Network inference methods}
Several methods have been proposed for inferring gene regulatory interactions from measured gene expression levels. Approaches employed include Bayesian networks, Boolean models, auto-regressive models, correlation-based models, clustering techniques and differential equation models, among others~\cite{gardner05,markowetz07,bansal07,nar1, marbach10}.
Some of them are static, while others take into account the dynamic aspects of the dependencies. Mutual information network inference methods are a class of network inference methods which infer regulatory interactions between genes based on pairwise mutual information. The low computational complexity and the low number of required samples are the main advantages of mutual information based inference methods. We have examined two commonly used state-of-the-art network inference methods based on pairwise mutual information: Algorithm for the Reconstruction of Accurate Cellular Networks, ARACNe~\cite{ARACNe2004} and Context Likelihood of Relatedness, CLR~\cite{CLR2007}.
ARACNe is based on an information theoretical approach that uses the concept of Mutual Information, MI, a measure of entropy, to determine the pairwise interaction between nodes by assessing the MI between them. It then applies a data processing inequality (DPI) to eliminate indirect interactions. The CLR algorithm is an extension of the network relevance approach. It is another information theoretic approach and computes the MI between two nodes, comparing it to the empirical background distributions of MI.
Regression based network inference methods comprise one of the largest network inference sub-categories, and we have studied 3 of the best regression based methods: GENIE3, TIGRESS and INFERELATOR. GENIE3 or GENIE~\cite{GENIE2010} decomposes the prediction of a regulatory network between $p$ genes into $p$ different regression problems such that in each, the expression pattern of one of the genes may be predicted from the expression patterns of all the other genes, using tree-based ensemble methods--Random Forests or Extra-Trees. TIGRESS~\cite{TIGRESS2012} formulates the inference problem as a sparse linear regression problem. It uses least angle regression (LARS) and adds an additional \emph{stability selection} criterion to assess the significance of
nodes in the regression. INFERELATOR uses regression and variable selection to identify transcriptional influences on genes based on the integration of genome annotation and expression data. In addition to these methods, we have used Correlation to reconstruct networks.
For Correlation, CLR, GENIE, INFERELATOR and TIGRESS, we used the implementations at the Michigan Institute of Technology's Broad Institute~\cite{GenePat2,GP-DreamOnline}. We used an ARACNe implementation in GE Workbench 2.5. 1~\cite{GEWB2.5.1}. Parameters are always default, set by GenePattern 2.0 or GE Workbench, unless otherwise stated.

\begin{table}[ht]
\label{T1}
{
\centering
\caption{Summary of assessed network inference models. }
\begin{tabular}{c|c|c}
\hline
\textbf{Method} &
\textbf{Category} & \textbf{Features}\\
\hline
ARACNe & Information- & MI estimated using a\\
& theoretic & copula-based approach \\
&& Use of the DPI to break \\
&& up fully connected triplets\\
\hline
CLR & Information-& MI dependencies(Gaussian\\
& theoretic & assumption)\\
&& Normalization of MI\\
\hline
GENIE3 & Tree-Based Methods & Decomposes into $p$ \\
&&different regression \\
&&problems. Prediction using \\
&&tree-based ensemble \\
&&methods\\
\hline
INFERELATOR & ordinary differential& Hybrid method\\ &Equations & involving differential \\
&& equations and Regression \\
\hline
TIGRESS & Regression Methods & Least Angle\\
&& Regression (LARS)\\
&& combined with\\
&&stability selection \\
\hline
Basic Correlation & Statistic & $-$ \tabularnewline
\hline
\end{tabular}
}
\end{table}

\subsection{The datasets}

Benchmarking is important in order to be able to understand the reliability of the reconstructed network. Traditionally, the assessment proceeds by collecting all curated interactions and considering them as true positives, while treating as false positives all predicted interactions between two genes that are not documented in the curated database. Such a method tends to overestimate the false-positive prediction rate while ignoring all new interactions. As a result, methods that merely reproduce current knowledge outperform those that do well at finding new results. To compensate for this, the gold standard networks were selected from among synthetic transcriptional networks proposed by Mendes et al.~\cite{mendes03}. These networks with well-defined topologies have been used by them to run \textit{in-silico} experiments simulating real laboratory micro-array experiments. They consist of 100 genes and are organized in an Erd\"{o}s-R\'enyi (ER), Smallworld (SW) or a scale-free (SF) topology. We have chosen 10 networks representing each topology: RND001 to 010, SW001 to 010 and SF001 to 010. The simulated data from these networks have been used as input for network inference methods.

\subsection{Graph-theoretic assessment method}

The performance of network inference methods has traditionally been evaluated using a confusion matrix with respect to the gold standard network, GSN, providing the number of true positives $TP$, true negatives $TN$, false positives $FP$ and false negatives $FN$. The measures in this confusion matrix have the following meaning in the context of this paper: TN refers to an edge that belongs neither to the predicted network nor to the gold standard network; FP is the number of predicted edges that do not belong to the gold standard network; FN is the number of edges in the
gold standard that are missing from the predicted network; and TP is the number of correct predictions of an edge in the gold
standard network.
To quantify network reconstruction performance, we first used receiver operator characteristic (ROC) and precision recall (PR) analysis.
We have used a threshold $\delta$ for discretization of edge values, where the weight $W_{i,j}$ for a particular edge is compared with $\delta$. If $|W_{i,j}|\leq \delta $, the edge $e_{i,j}$ is assumed to be present in the network $e_{i,j}\in E$, and absent otherwise. The resulting network with edge set $E$ is then compared
against the gold standard network, and sensitivity, specificity
and precision are computed for a given $\delta$. This is then repeated by varying $\delta$, and sensitivity is plotted over specificity for different values of $\delta$ in a ROC plot. Finally, the ROC curve can be summarized by computing the area under the curve~\cite{nar1}.
As our second approach, we have used the Jaccard coefficient~\cite{Jaccard}. This commonly used similarity metric measures the probability that the two networks, the gold standard and the inferred network, have common edges, focusing on randomly selected edges in either of the networks.
\begin{center}
JaccardCoefficient(GSN,IN)$ := |E(GSN)\bigcap E(IN)|/|E(GSN)\bigcup E(IN)|. $
\end{center}

\subsubsection{Topological Indices Enrichment Analysis \\}

We have compared the topological indices by borrowing ideas from gene set enrichment analysis (GSEA). We call our procedure Topological Indices Enrichment Analysis, TIEA. GSEA is one of the most widely used methods for detecting differentially expressed gene sets. GSEA~\cite{bet} is a discrete version of the weighted Kolmogorov-Smirnov test, which is applied to a running sum statistic over ranked lists, counting how often elements are or are not in the list of interest. Unlike in the analysis of gene expression data, the sets here were defined not by genes but by nodes from networks, and ranked not based on expression but on the topological index of interest.
For TIEA, nodes are first ranked by topological score. Then a "running sum" statistic is calculated for each network, based on the ranks of subsets of nodes in the network, relative to those of non-members. An enrichment score (ES) is defined to be the maximum of the running sum across all nodes, which corresponds to a weighted Kolomogorov-Smirnov statistic. The equation for the calculation of ES for the sorted list was processed from top to bottom, and two running sums, $RS_{N_k}$ and $\overline{RS}_{N_k}$, were computed. $RS_{N_k}$ was increased by one each time a node belonged to $N_k$, and $\overline{RS}_{N_k}$ each time a node belonged to the complementary set $N_k$:
\begin{center}
$RS_{N_k}(i)=\Sigma_{\substack{n_j \in N_k \\ j\leq i}}\, \dfrac{1}{\Vert N_{k} \Vert}$
\end{center}
\begin{center}
$\overline{RS}_{N_k}(i)=\Sigma_{\substack{n_j \notin N_k \\ j\leq i}}\, \dfrac{1}{\Vert N - N_{k} \Vert}$
\end{center}
\begin{center}
$TIES_{N_k}=max(\vert RS_{N_k} - \overline{RS}_{N_k}\vert)$
\end{center}

\subsubsection{Topological indices}

A network, graph $G$, consists of a set of nodes representing biological entities $V(G)$, while the edges $E(G)$ denote relationships between node pairs. Its topological structure is the most basic and direct information available about a network. The architectural features of biological networks can be roughly categorized into three classes: individual, local and global features. Individual features are topological properties associated with only one node, including degree and centrality measures; global features involve all the vertices in networks, while local features are those behaviors that involve part of the network rather than the whole network containing motifs~\cite{alon06} and communities. This paper confines itself to individual features.
We focus on the preservation of diameter, average path length, clustering, centrality and degree distributions.

\begin{defn}
The diameter of a network is the largest distance between any two
nodes in the network
\end{defn}
\begin{defn}
The average path length is the average distance between any two
nodes in the network.
\end{defn}
Average path length is bounded from above by the diameter; in some
cases it can be much shorter than the diameter. If the network is not connected, one often checks the diameter and the average path length in the largest component.
\begin{defn}
The overall clustering coefficient Cl(G) is given by \\
\begin{center}
$Cl(g) = \dfrac{3 \, *\, number\, of\, triangles\, in\, the \,network}{number\, of\, connected\, triples\, of\, nodes}$ \\
\end{center}
where a ``connected triple" refers to a node with edges linked to an
unordered pair of nodes.
\end{defn}
\begin{defn}
The individual clustering for a node i is
\begin{center}
$Cl_i(g) =\dfrac{number\, of\, triangles\, connected\, to\, vertex \,i}{number\, of\, triples\, centered\, at \,i} $
\end{center}
\end{defn}

\subsection{Information-theoretic assessment method}

\subsubsection{Classical information theory}

Central to information theory is the concept of Shannon's information entropy~\cite{shannon}, which quantifies the average number of bits needed to store or communicate a message. Shannon's entropy determines that one cannot store (and therefore communicate) a symbol with $n$ different symbols in less than $\log(n)$ bits. Shannon's entropy determines a lower limit below which no message can be further compressed, not even in principle. For an ensemble $X(R, p(x_i))$, where $R$ is the set of possible outcomes (the random variable), $n=|R|$ and $p(x_i)$ is the probability of an outcome in $R$. The Shannon information content or entropy of $X$ is then given by

$$H(X)=-\sum_{i=1}^n p(x_i) \log_2 p(x_i)$$

The Shannon Entropy (or simply entropy) of a graph $G$ is simply defined by $H(A(G))=-\sum_{i=1}^n P(A(x_i)) \log_2 P(A(x_i))$, where $G$ is the random variable with $n$ possible outcomes (all possible adjacency matrices of size $|V(G)|$). For example, a completely disconnected graph $G$ with all adjacency matrix entries equal to zero has entropy $H(A(G))=0$ (the same as the complete one with all self-loops), because the number of different symbols in the adjacency matrix is 1 (of a possible total of 2). However, if a different number of 1s and 0s occur in $A(G)$, then $H(A(G))\neq0$.

\subsubsection{Algorithmic complexity}

A more powerful measure of information content and randomness than Shannon entropy, with which to find not only statistical but recursive regularities, is the concept of Kolmogorov complexity~\cite{kolmo,chaitin}---denoted by $K$.The Kolmogorov complexity of a string $s$ is given by:

$$K(s)=\min\{|p| : U(p)=s\}$$

\noindent that is, the length (in bits) of the shortest program $p$ that when running on a universal Turing machine $U$ outputs $s$ upon halting.

Despite its great power, $K$ comes burdened with a technical inconvenience, its \textit{semi-computable}) nature, i.e. no effective algorithm exists which takes a string $s$ as input and produces the exact integer $K(s)$ as output~\cite{kolmo,chaitin}. This is related to a common problem in computer science known as the undecidability of the halting problem~\cite{turing}---referring to the inability to know whether or not a computation will eventually stop.

Kolmogorov complexity can be understood in terms of uncompressibility. If an object, such as a biological network, is highly compressible, then $K$ is small and the object is said to be non-random. However, if the object is uncompressible then it is considered algorithmically random. Despite the inconvenience, $K$ can be effectively approximated by using, for example, compression algorithms (here we have used \textit{Compress}).

\subsubsection{Algorithmic probability}

There is another seminal concept in the theory of algorithmic information~\cite{solomonoff,levin}, namely the concept of \textit{algorithmic probability}. The algorithmic probability of a string $s$ gives the probability that a valid random program $p$ written in bits uniformly distributed produces the string $s$ when run on a universal (prefix-free~\footnote{The group of valid programs forms a prefix-free set (no element is a prefix of any other, a property necessary to keep $0 < m(s) < 1$).}) Turing machine $U$. In equation form this can be rendered as:

$$m(s) = \sum_{p:U(p) = s} 1/2^{|p|}$$

That is, the sum over all the programs $p$ for which $U$ outputs $s$ and halts.

The algorithmic probability measure $m(s)$ is related to Kolmogorov complexity $K(s)$ in that $m(s)$ is at least the maximum term in the summation of programs, given that the shortest program carries the greatest weight in the sum. The algorithmic Coding Theorem~\cite{levin} further establishes the connection between $m(s)$ and $K(s)$ as follows:

$$|-\log_2 m(s) - K(s)| < \mathcal{O}(1)$$

\noindent where $\mathcal{O}(1)$ is an additive value independent of $s$. Hence:

$$K_m(s)=-\log_2 m(s) + \mathcal{O}(1)$$

One can see then that it is possible to approximate $K$ by approximating $m$ (hence the notation $K_m$), with the added advantage that $m(s)$ is more sensitive to small objects~\cite{ctm} than the traditional approach to $K$ using lossless compression algorithms, which typically perform poorly where small objects are concerned (e.g. small graphs).

\subsubsection{Loss/gain of algorithmic information}

The approach to determining the algorithmic complexity of a graph thus involves considering how often the adjacency matrix of a motif is generated by a random Turing machine on a 2-dimensional array~\cite{bdm}. This is called the\textit{Block Decomposition Method} (BDM)~\cite{bdm} as it requires the partition of the adjacency matrix of a graph into smaller matrices, using which we can numerically calculate its algorithmic probability by running a large set of small 2-dimensional deterministic Turing machines, and then, by applying the algorithmic Coding theorem, its Kolmogorov complexity. See the Supp. Inf. for further details.

We will say that a method has lost $c$ information if $C(N)+c=C(G)$, or has introduced spurious information if $C(N)-c=C(G)$, where $G$ is the true information content of the original (e.g. gold standard) network, $N$ is the reconstructed network and $C$ is the evaluation algorithm used to assess the information content of the networks (e.g. Shannon entropy, lossless compression or algorithmic complexity via BDM). Notice that $C(N)=C(G)$ does not imply that $N$ has recovered all the properties of $G$ (in the unlikely case of exact equality).

\section{Results}

ARACNe, Basic Correlation and GENIE3 successfully inferred networks from inputs; CRL returned empty networks for all inputs; INFERELATOR broke down due to ``zero variance'' for the subset of SF networks, but worked for the other sets; TIGRESS returned results only for seven networks in the ER subset.

\begin{figure}[!h]
\centering\includegraphics[width=4.8in]{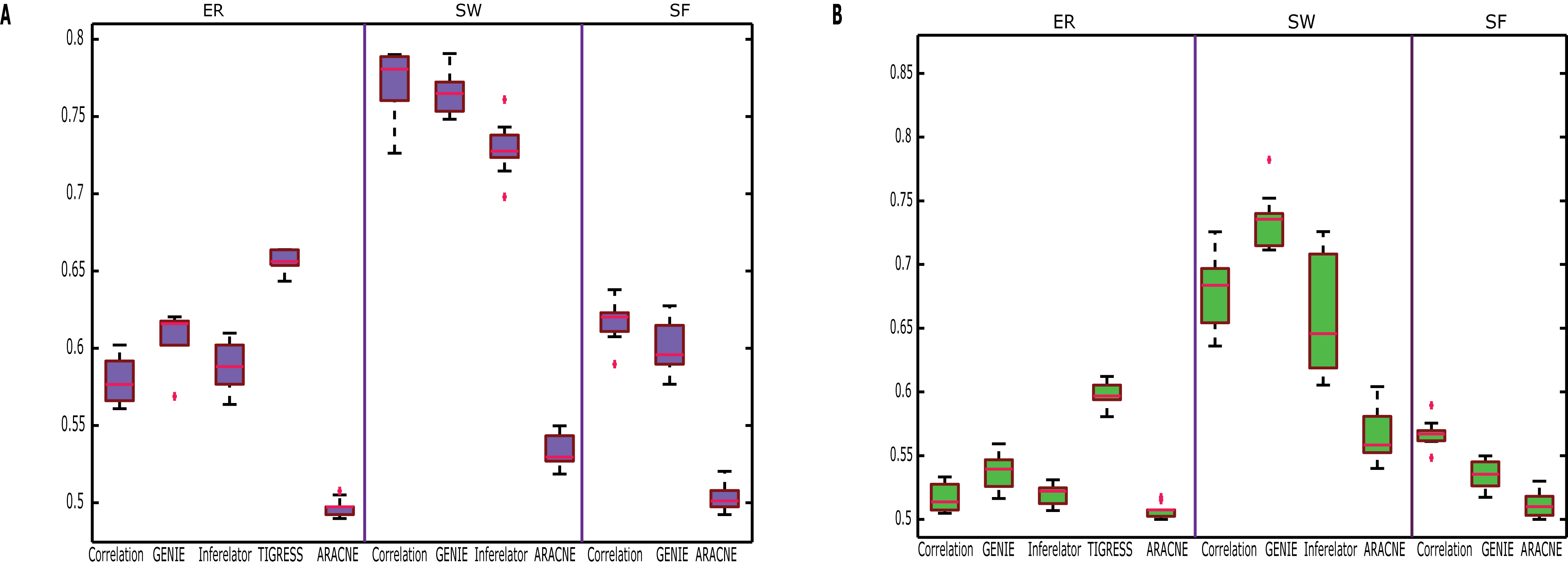}
\centering\caption{Evaluation of simulated data. Network reconstruction was performed for all networks in all 3 categories. Shown are \textbf{A:} the distribution of the area under the ROC curve (AUCROC), and \textbf{B:} the area under the precision recall curve (AUCPR) of 30 sets of simulated data, over the different topologies. From left to right: Correlation, GENIE, Inferelator, TIGRESS and ARACNe. AUC $= 0.5$ is the expected result for random guessing.}
\label{fig_2}
\end{figure}

The AUCROC values for each algorithm and dataset can be seen in \ref{fig_2}A. All models turn in performances markedly superior to random guesswork, except ARACNe. Despite the significance of the difference, its magnitude, compared against random guessing is not, in the best case, more than $10\%$. Relative to the datasets ER and SF, the best performer is GENIE3 (or simply GENIE), while for SW the best performer is Correlation, but there is no significant difference between GENIE3 and Correlation for SW networks. When we examine the area under the precision recall curve (AUCPR) of the models for the same set of networks, we find that GENIE3 significantly outperforms Correlation (see \ref{fig_2}B). This shows the higher number of false positives in the network predicted by Correlation.

\begin{figure}[!h]
\centering\includegraphics[width=4.8in]{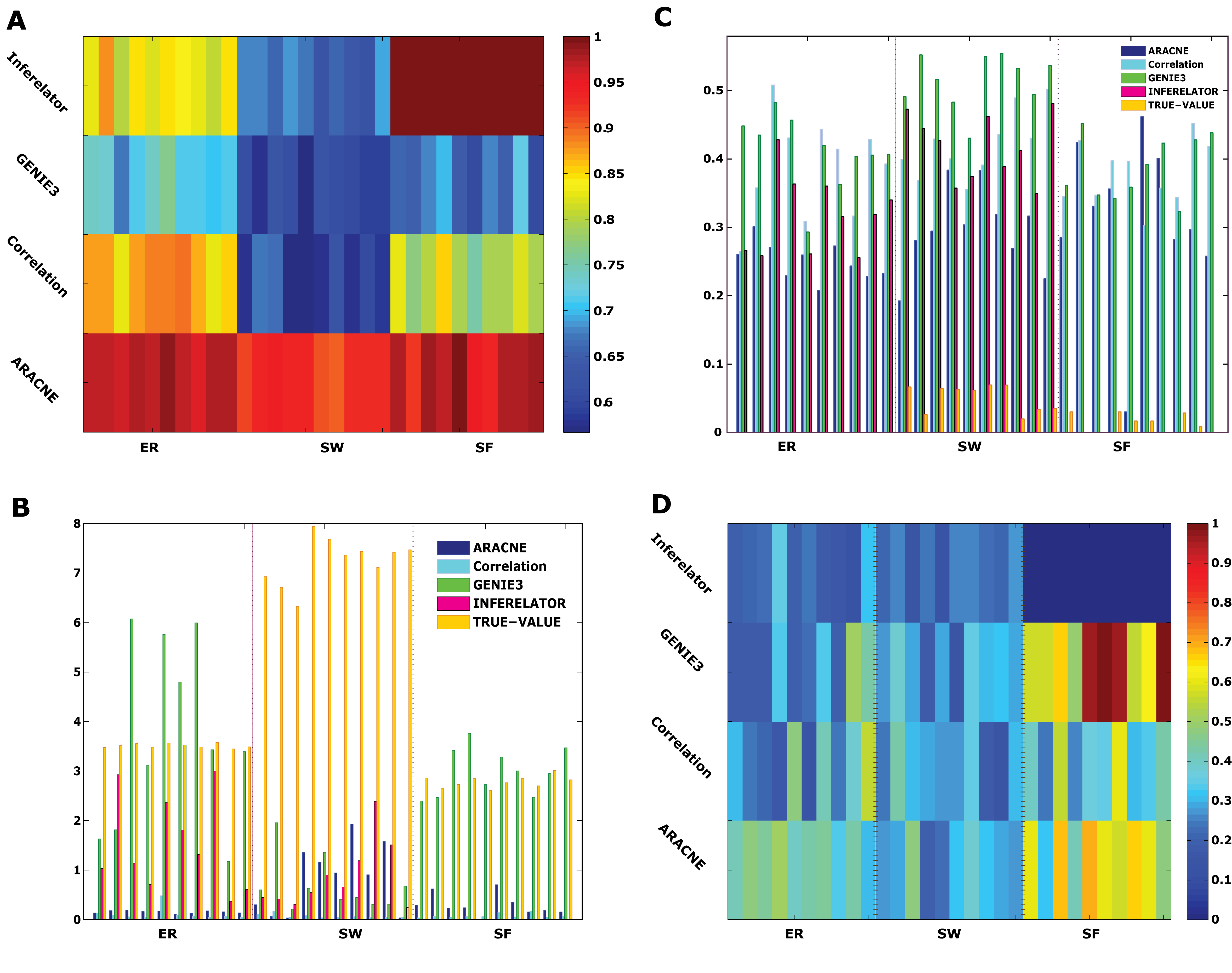}
\centering\caption{Performance of all models based on different measures and assessment approaches. Assessment of models' performance using Jaccard distance (panel A), Average shortest path (panel B), Clustering coefficient (panel C). The hub enrichment score is shown in panel D.}
\label{fig_1}
\end{figure}

It has been shown that most biological networks are scale-free networks, yet all methods perform significantly better for SW topology. All in all, ARACNe turns in the worst performance on all network categories. This may be due to the number of arbitrary parameters and the effect of the cutoff values for the network reconstruction, which was set to the value that produced the closest number to the number of edges in the gold standard networks (200 edges). Surprisingly, Basic Correlation turns in a performance comparable to other methods in all categories when statistical measures are the parameters being compared.

Then we assessed the Jaccard distance between each GSN and its corresponding inferred network. Figure \ref{fig_1}A shows the result. As can be seen, the overall picture remains the same as for AUCROC, the topology of the underlying network significantly affects the performance of all methods, and GENIE3 outperforms other methods. Then we have used topological indices to assess all inference methods. We have done this using both enrichment scores and Euclidean distance. An overview of the results has been shown in \ref{T2} and in Figures \ref{fig_1}B and C. We found discrepancies in the rankings obtained using the 3 approaches. As has been mentioned, ARACNe was the worst performer, but when assessed using a topological index, it was one of the two most effective methods. To take another example, GENIE3 outperforms ARACNe at prediction, and it is closest to the GSN if we focus on the shortest path and measure the difference using Euclidean distance, but ARACNe is better when the clustering coefficient is the parameter being compared.

\begin{table}
\centering
\begin{tabular}{c|c|c|c|c}
\hline
& \textbf{Jacc. dist} & ${\Delta CC}$ & ${\Delta SP}$ & ${\Delta Diam}$\tabularnewline
\hline
\textbf{ARACNe}$^{a}$ & 0.9550 & -0.1547 & 4.088 & 3.50\tabularnewline
\hline
\textbf{TIGRESS}$^{c}$ & 0.6225 & -0.1562 & -0.844 & -6.14\tabularnewline
\hline
\textbf{Basic Correlation}$^{a}$ & 0.7567 & -0.2634 & 4.428 & 4.37\tabularnewline
\hline
\textbf{GENIE3}$^{a}$ & 0. 6607 & -0.3056 & 2.045 & -3.93\tabularnewline
\hline
\textbf{INFERELATOR}$^{b}$ & 0.7477 & -0.3350 & 4.183 & 0.95\tabularnewline
\hline
\end{tabular}
\caption{\label{T2}The Jaccard distance, clustering coefficient difference(${\Delta CC}$), shortest path difference (${\Delta SP}$) and diameter difference (${\Delta Diam}$) between all GSN and inferred networks have been calculated for all models using Euclidean distance. Averages over all 30 networks are shown. All networks, $^{a}$: Based on 30 networks, $^{b}$: based on 20
networks, $^{c}$: based on 7 RND networks.}
\end{table}

We then compared all models using our new approach--TIEA. First, we considered hub enrichment scores for all models. For this purpose we selected $10\%$ of the top hub nodes in the GSN and calculated the enrichment score of this set in all prediction models. Of all the methods, we found ARACNe to have the highest median hub enrichment score. GENIE3 got the highest score for SF networks; see \ref{fig_1}D. We used the same procedure for the clustering coefficient and diameter of a network. ARACNe's ranking improves dramatically when we use TIEA for evaluation. This shows the power of ARACNe when it comes to capturing the central section of a network. On the other hand, GENEI fares better at predicting the totality of a network. TIEA measures the ability of a model to predict the most important features of a network, such as Euclidean distance, by giving equal weight to all nodes. The result has been reported in tables \ref{T2} and \ref{T3}. A comparison of Tables \ref{T2} and \ref{T3} brings home the importance of picking a model suited to the specific task at hand.

\begin{table}
\centering
\begin{tabular}{c|c|c|c|c}
\hline
& \textbf{HubES} & \textbf{CCES} & \textbf{SPES} & \textbf{DiamES} \tabularnewline
\hline
\textbf{ARACNe}$^{a}$ & 0.4336 & 0.4847 & 0.3429& 0.4873 \tabularnewline
\hline
\textbf{GENIE3}$^{a}$ & 0.4277 & 0.4305 & 0.3587 & 0.4902 \tabularnewline
\hline
\textbf{INFERELATOR}$^{b}$ & 0.2187 & 0.4132 & 0.5362 & 0.1393\tabularnewline
\hline
\textbf{Basic Correlation}$^{a}$ & 0.3252 & 0.1342 & 0.2248 & 0.2042\tabularnewline
\hline
\textbf{TIGRESS}$^{c}$& 0.1623 & 0.1230 & 0.1181 & 0.2362 \tabularnewline
\hline
\end{tabular}
\caption{\label{T3}The Hub enrichment score (HubES), clustering coefficient enrichment score (CCES), shortest path enrichment score (SPES) and diameter enrichment score (DiamES) are shown. Values shown are average values over 30 networks. All networks, ${a}$: Based on 30 networks, ${b}$: based on 20 networks, ${c}$: based on 7 RND networks. }

\end{table}

\begin{figure}[!h]
\centering
\includegraphics[width=4.8in]{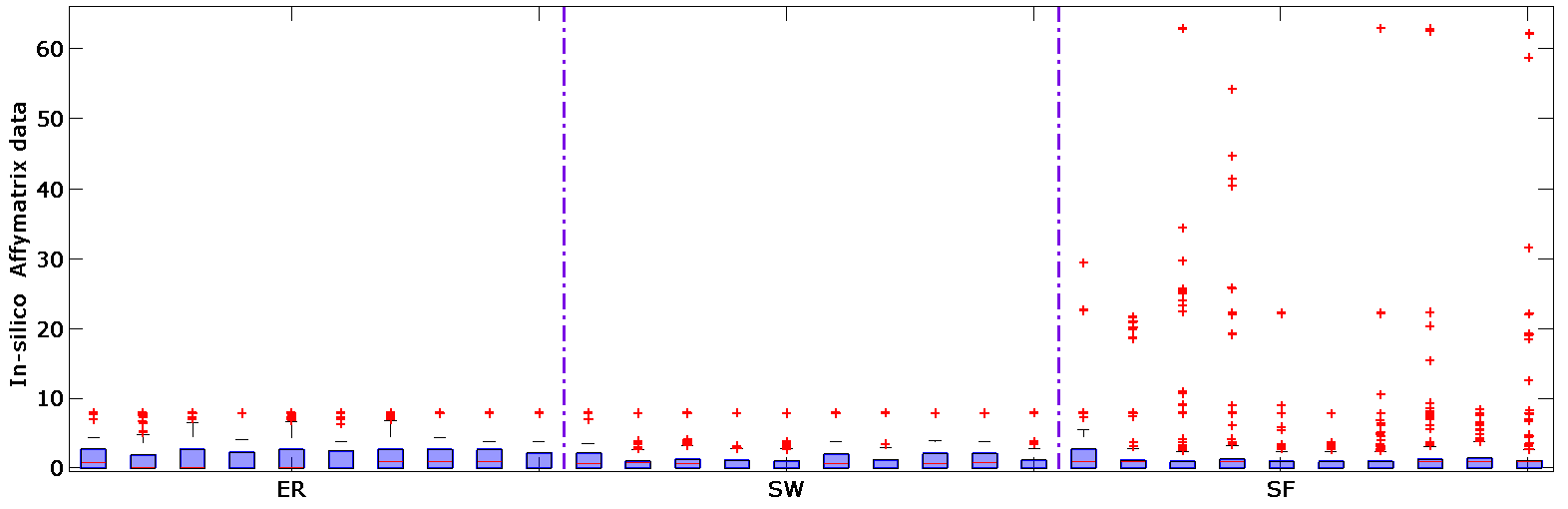}
\centering\caption{Shown is the distribution of the 30 input sets of simulated data that has been used for the network inference task generated from the gold standard network with the different topologies.}
\label{fig_3}
\end{figure}

The significant differences between the methods based on topology motivate us to study the inputs with a view to finding indices that would guide the choice of different topological measures. Affymetrix was used in the 30 in-silico networks to be inferred. As can be seen in~\ref{fig_3}, there is no significant difference in the mean values of the input generated from ER and SF networks, but the input generated from SW networks has significantly lower mean values (paired T test). While there is no significant difference between the mean values of input generated from ER and SF networks, input generated from SF networks are distinguishable by the number of outliers. Then we looked at the coefficient of variation (CV) for all 3 types of inputs. Input generated from SF networks has the highest CV (7.895), followed by input generated from SW networks (2.896) and finally input generated from ER networks (1.989).

\subsection{Method under- and over-fitting}

\begin{figure}[!h]
\centering
\hspace{.5cm}Adj Entropy \hspace{4cm} \textit{ } Degree Entropy\\
\includegraphics[width=2.2in]{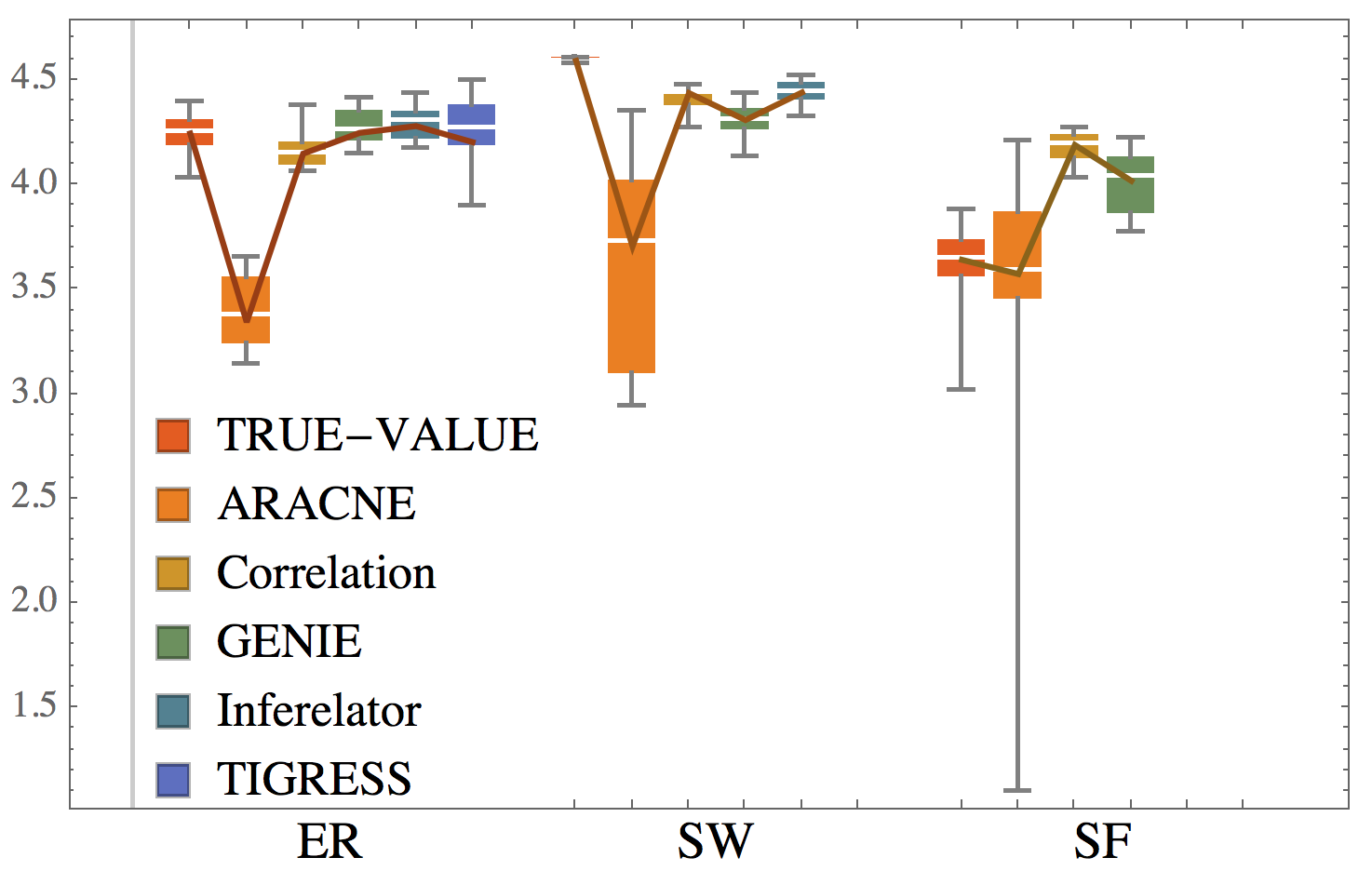}\hspace{.3in}\includegraphics[width=2.2in]{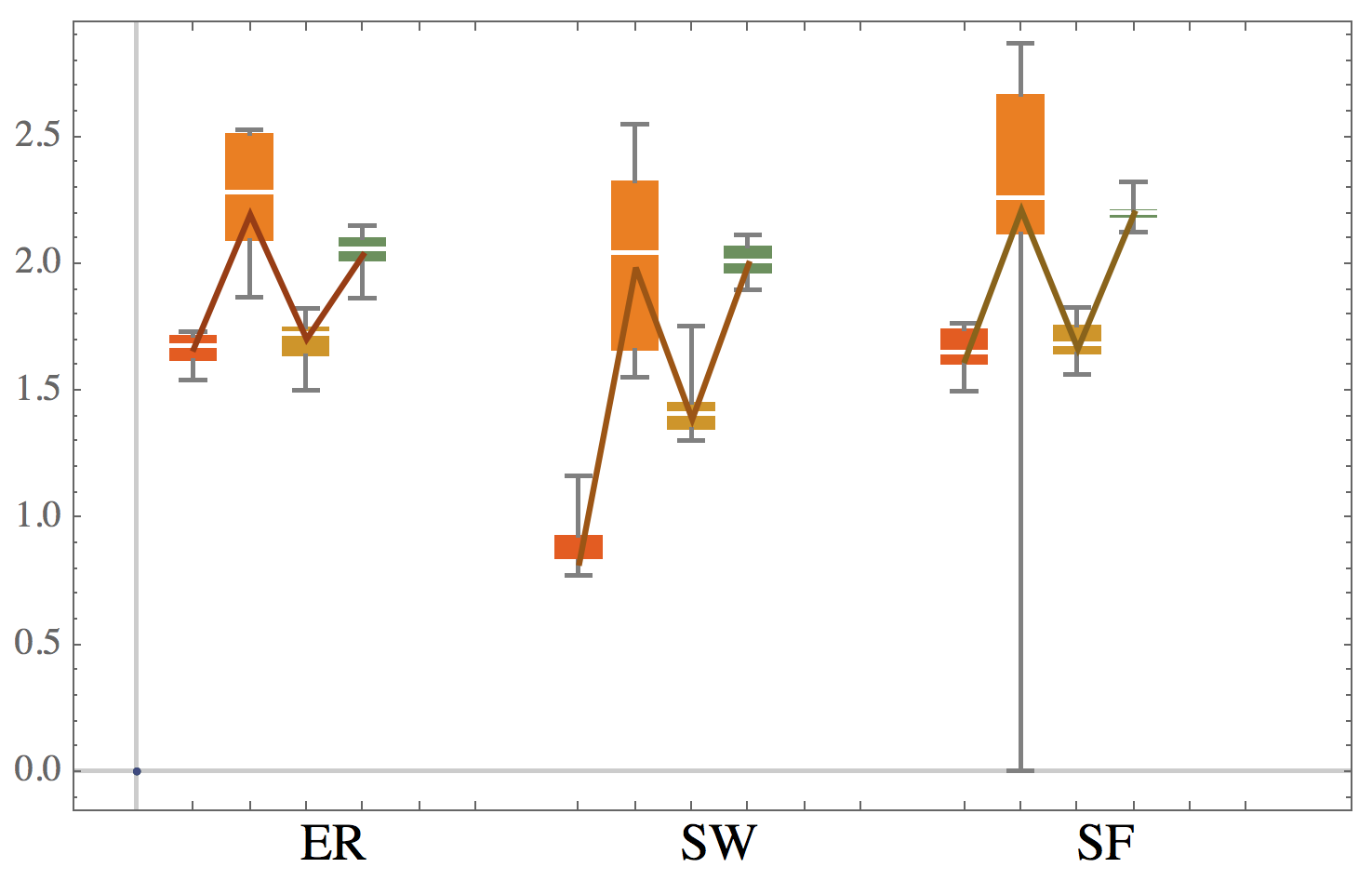}\\
\medskip

\hspace{.5cm} BDM \hspace{4.5cm} \textit{ } Compress
\includegraphics[width=2.2in]{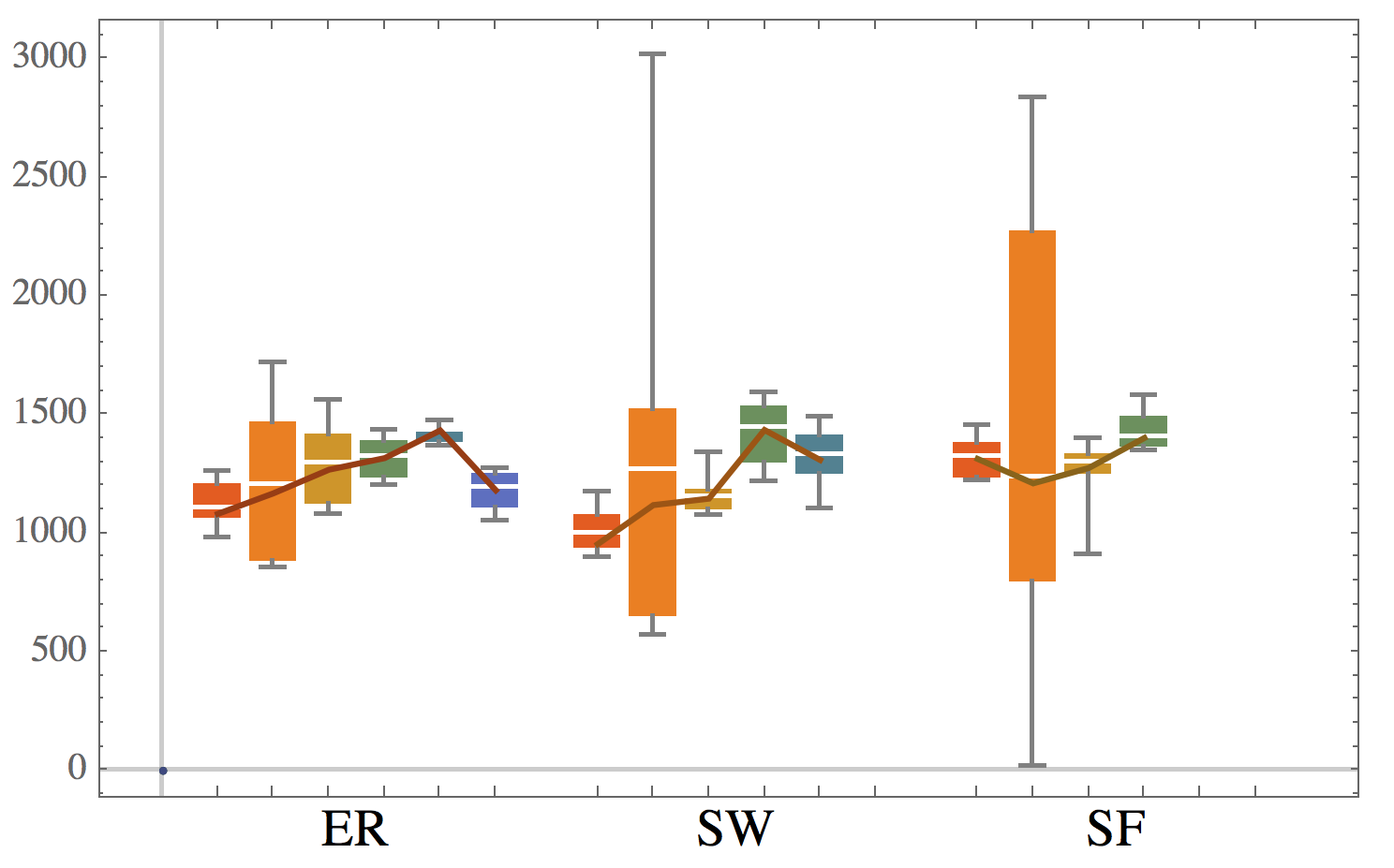}\hspace{.3in}\includegraphics[width=2.2in]{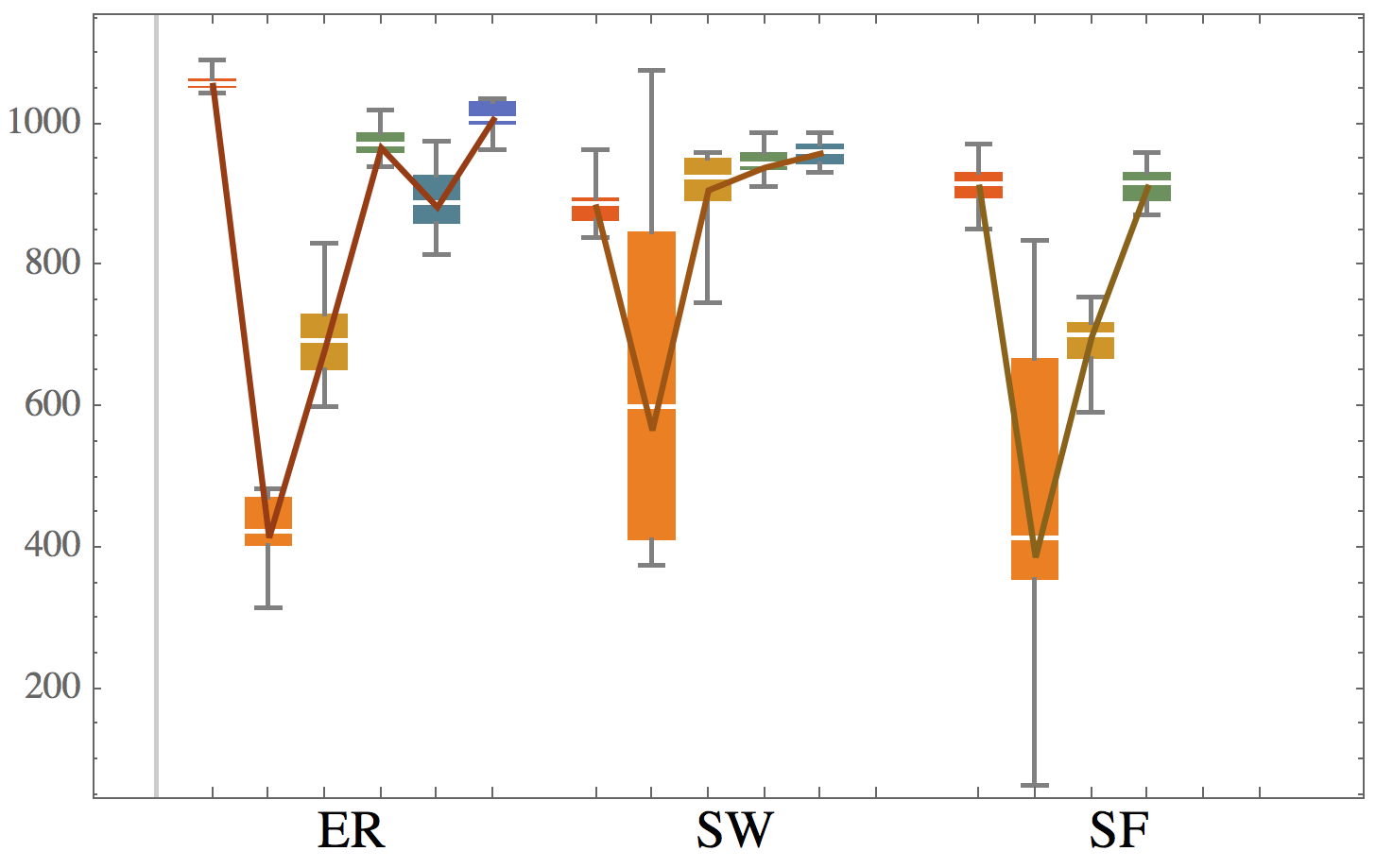}
\centering\caption{Loss and preservation of information content in networks reconstructed from simulated data by different methods, as compared to the information content of the original network.}
\label{fig_4}
\end{figure}

Concepts and tools from information theory and algorithmic complexity can be drawn upon to perform a more general analysis of network profiling~\cite{zenilcomplexnetworks,zenilphysicaa}. Here we applied these tools in the evaluation of network reconstruction methods. Unlike graph-theoretic or entropic measures, algorithmic complexity tools (particularly based on algorithmic information content such as BDM) are in a strong sense ``parameter-free'' because there is no need to focus on a particular local or global graph-theoretic property to evaluate but rather on the amount of information lost or the spurious complexity introduced in the network reconstruction task. We have used four different measures: Shannon entropy as applied to the adjacency matrix and the degree sequence, algorithmic complexity (by means of approximations with BDM~\cite{ctm,bdm}) and compression (Compress). The results are depicted in Figure~\ref{fig_4}. The plot shows how much information is lost when reconstructing a network (True-value) with 5 different algorithms for different topologies.

The first point worth noting is the significant differences among topologies and methods in their complexity values for both gold standard networks, reconstructed networks and their relative order. While for Shannon entropy on adjacency matrices, SW followed by ER networks have the greatest classical information content, for both BDM and Compress they have a lower complexity or algorithmic information content than SF networks. This can be understood in terms of the fact that synthetic ER networks are actually recursive and not algorithmically random, as are SW networks, while appearing random to Shannon entropy. However, SF networks are the ones with greatest algorithmic complexity according to BDM approximations and lossless compression. In all cases ARACNe is an outlier with the greatest variance and the least significance across topologies hence the least discriminant but also the one that can either show the greatest loss of information or introduce spurious information. According to BDM, Correlation is the closest method preserving the information content of the network but GENIE3 and Inferelator alternate between ER and SW.

\section{Conclusions}

Current efforts aim to understand the individual strengths and weaknesses of various network reconstruction methods by applying them to equal and different data sets. Generally, sensitivity, specificity, precision and the (ROC) curve are calculated to illustrate the performance of a particular approach. However, there are more aspects of network construction that should be taken into consideration. Here we suggest using topological indices to evaluate network inferences, and furthermore we introduce a concept similar to gene set enrichment for network inference evolution. Depending on the ultimate goal, one can use Euclidean distances or TIEA to compare models.

Our new assessment process revealed a new feature of the models: we observed that the performance of a model depended on the topological index. We have introduced a graph and information-theoretical perspective on the problem that allows one to study particular substructures of networks and not be limited to studying networks in their entirety. In this paper, several commonly used computational approaches for constructing gene regulatory networks are compared, using both topological and statistical indices.

We have used data produced by synthetic networks to address questions such as the following: Which topologies are best suited for a specific network inference method? Which models work best for predicting specific properties of a network? Obviously, synthetic data cannot reflect the complexity of a real biological system. However, standards are still unavailable for evaluating different inference methods using real biological data. Results obtained using three different datasets show that the overall performance of the models assessed is poor.

There are significant differences in the results obtained with the datasets (ER, SW and SF). In general, we observe that when based on statistical measures, network inference methods could be said to perform better with small world networks, while when based on topological measures, they perform better with SF networks.

A few conclusions can be drawn from this exercise. First, GENIE3 and INFERELATOR performed well in constructing the global network, while ARACNe did well in identifying a few connections with high specificity. Surprisingly, Correlation performed well in constructing the global network, doing better than ARACNe and approaching GENIE3. GENIE3 performed well in both respects, but it is not suitable for identifying the hub nodes which can often be of biological interest. ARACNe performed well in identifying the hub genes.

We have shown that the inputs generated from these networks have different statistical and information-theoretic signatures. For example, inputs generated from SF networks are easily distinguishable by their very high variance with ARACNe displaying the greatest variance for information-theoretic indices. This information can be used in choosing a proper method for inference, specially when anything is known about the topology of the source. The general literature favors, for example, SF networks when it comes to biological and other kind of networks.

However, while most network inference methods may have been designed to better reconstruct networks following a scale-free (SF) distribution. However, networks may or may not follow a scale-free distribution, the only certainty is that they distribute somewhere between random and trivial networks because they tend to encode important non-trivial information.

We have the same situation than for the graph-theoretic case when it comes to the general information loss and introduction of random complexity in a reconstructed network, that no single method outperforms all others when applied to all topological cases. However, we found ARACNe is more prone to under- and over-fit the complexity than that contained in the source network. Therefore, since there is no single method that outperforms other methods in all respects, care should be taken to choose an appropriate method to suit the purpose of the study.

\section*{Acknowledgments}

This work was supported in part by the Foundational Questions Institute (HZ), the VINNOVA (VINNMER) Marie-Curie Fellowship, (NK), StratNeuro (JT,NK), AFA Insurance(JT), Torsten S{\"o}derberg Foundation (JT), STATegra (JT, HZ), the Swedish Research Council - Vetenskapsr\r{a}det (HZ), the Stockholm County Council and the Swedish Research Council. The funders played no role in the design of the study, in data collection and analysis, in the decision to publish, or in the preparation of the manuscript.


%

\newpage

\section*{Supplemental Information}

\subsection{Network reconstruction}

Tables~\ref{t1}, \ref{t2} and \ref{t3} show the edge count for all 30 reconstructed networks. ARACNe was the only method returning a significantly greater number of high confidence edges, displaying the least sensitivity among all the methods by assigning the highest confidence to 50\% of the spurious edges, as compared to the true value of 200 edges coming from the gold standard (the synthetic data in the Mendes database). In all other cases, the cutoff value of each of the methods was chosen so that there were around the true number of predicted edges.

\begin{table}[ht]
\centering
\begin{tabular}{c|c|c|c|c}
\hline
\textbf{ARACNe} & \textbf{Basic correlation} & \textbf{GENIE3} & \textbf{Inferelator} & \textbf{TIGRESS}\\
\hline
282 & 202 & 200 & 200 & 201 \\
377 & 202 & 199 & 200 & 0 \\
240 & 202 & 200 & 200 & 0 \\
258 & 200 & 200 & 199 & 199 \\
260 & 196 & 200 & 200 & 201 \\
209 & 200 & 200 & 202 & 0 \\
287 & 196 & 200 & 199 & 201 \\
250 & 206 & 201 & 202 & 200 \\
260 & 200 & 199 & 201 & 0 \\
254 & 200 & 201 & 198 & 200 \\
\hline
\end{tabular}
\caption{\label{t1}Edge count for reconstructed ER networks.}
\end{table}

\begin{table}[ht]
\centering
\begin{tabular}{c|c|c|c|c}
\hline
\textbf{ARACNe} & \textbf{Basic correlation} & \textbf{GENIE3} & \textbf{Inferelator} & \textbf{TIGRESS}\\
\hline
72 & 202 & 200 & 200 & 0 \\
79 & 200 & 200 & 200 & 0 \\
85 & 200 & 200 & 201 & 0 \\
143 & 200 & 200 & 203 & 0 \\
124 & 200 & 200 & 197 & 0 \\
180 & 200 & 200 & 200 & 0 \\
136 & 200 & 200 & 200 & 0 \\
220 & 202 & 200 & 200 & 0 \\
256 & 200 & 200 & 200 & 0 \\
59 & 202 & 200 & 202 & 0 \\
\hline
\end{tabular}
\caption{\label{t2}Edge count for reconstructed SW networks. Null values mean that the method did not produce any edges.}
\end{table}

\begin{table}[ht]
\centering
\begin{tabular}{c|c|c|c|c}
\hline
\textbf{ARACNe} & \textbf{Basic correlation} & \textbf{GENIE3} & \textbf{Inferelator} & \textbf{TIGRESS}\\
\hline
194 & 200 & 200 & - & - \\
442 & 200 & 200 & - & - \\
262 & 198 & 200 & - & - \\
394 & 200 & 200 & - & - \\
3 & 218 & 200 & - & - \\
612 & 200 & 200 & - & - \\
396 & 208 & 200 & - & - \\
158 & 202 & 200 & - & - \\
298 & 202 & 200 & - & - \\
83 & 200 & 200 & - & - \\
\hline
\end{tabular}
\caption{\label{t3}Edge count for reconstructed SF networks - means that the method did not produce any networks.}
\end{table}

Tables~\ref{bt1}, \ref{bt2} and \ref{bt3} show the cutoff values. Cutoff values are normalized ranging from 0 (lowest confidence) to 1 (highest confidence). The strongest cutoff value for ARACNe returned about 50\% more spurious edges than the true value (200); for all others a value between 0 and 1 was chosen so that the number of predicted edges was about the number of actual true positive edges (200).

\begin{table}[ht]
\centering
\begin{tabular}{c|c|c|c|c}
\hline
\textbf{ARACNe} & \textbf{Basic correlation} & \textbf{GENIE3} & \textbf{Inferelator} & \textbf{TIGRESS}\\
\hline
1 & 0.794 & 0.0598 & 0.429 & 0.230 \\
1 & 0.744 & 0.0568 & 0.408 & - \\
1 & 0.728 & 0.0569 & 0.453 & - \\
1 & 0.748 & 0.0516 & 0.423 & 0.219 \\
1 & 0.696 & 0.0463 & 0.411 & 0.234 \\
1 & 0.682 & 0.0525 & 0.436 & - \\
1 & 0.699 & 0.0512 & 0.363 & 0.244 \\
1 & 0.772 & 0.0556 & 0.513 & 0.229 \\
1 & 0.716 & 0.0517 & 0.436 & - \\
1 & 0.770 & 0.0578 & 0.477 & 0.238 \\
\hline
\end{tabular}
\caption{\label{bt1}Cutoff values for the ER graphs - means that no edges or networks (not even output) were produced, for reasons having to do with the methods themselves.}
\end{table}

\begin{table}[ht]
\centering
\begin{tabular}{c|c|c|c|c}
\hline
\textbf{ARACNe} & \textbf{Basic correlation} & \textbf{GENIE3} & \textbf{Inferelator} & \textbf{TIGRESS}\\
\hline
1 & 0.688 & 0.0784 & 0.409 & - \\
1 & 0.770 & 0.0705 & 0.465 & - \\
1 & 0.754 & 0.0721 & 0.510 & - \\
1 & 0.653 & 0.0798 & 0.372 & - \\
1 & 0.722 & 0.0782 & 0.449 & - \\
1 & 0.688 & 0.0813 & 0.440 & - \\
1 & 0.697 & 0.0838 & 0.498 & - \\
1 & 0.641 & 0.0780 & 0.421 & - \\
1 & 0.688 & 0.0797 & 0.474 & - \\
1 & 0.683 & 0.0767 & 0.371 & - \\
\hline
\end{tabular}
\caption{\label{bt2}Cutoff values for the SW graphs. - means that no edges or networks (not even output) were produced, for reasons having to do with the methods themselves.}
\end{table}

\begin{table}[ht]
\centering
\begin{tabular}{c|c|c|c|c}
\hline
\textbf{ARACNe} & \textbf{Basic correlation} & \textbf{GENIE3} & \textbf{Inferelator} & \textbf{TIGRESS}\\
\hline
1 & 0.672 & 0.0607 & - & - \\
1 & 0.524 & 0.0610 & - & - \\
1 & 0.699 & 0.0544 & - & - \\
1 & 0.577 & 0.0532 & - & - \\
1 & 0.531 & 0.0487 & - & - \\
1 & 0.415 & 0.0569 & - & - \\
1 & 0.632 & 0.0576 & - & - \\
1 & 0.460 & 0.0521 & - & - \\
1 & 0.702 & 0.0516 & - & - \\
1 & 0.565 & 0.0553 & - & - \\
\hline
\end{tabular}
\caption{\label{bt3}Cutoff values for the SF graphs - means that no edges or networks (not even output) were produced, for reasons having to do with the methods themselves.}
\end{table}

\subsection{Calculation using the Block Decomposition method}

The overall complexity of the original adjacency matrix is the sum of the complexity of its parts, albeit with a logarithmic penalization for repetitions, given that $n$ repetitions of the same object only adds $\log n$ to its overall complexity, as one can simply describe a repetition in terms of the multiplicity of the first occurrence. More formally, the Kolmogorov complexity of a labeled graph $G$ obtained by means of the $BDM$ is defined as follows~\cite{bdm}:

\begin{equation}
\label{newecaeq}
BDM(G,d) = \sum_{(r_u,n_u)\in A(G)_{d\times d}} \log_2(n_u)+K_m(r_u)
\end{equation}
where $K_m(r_u)$ is the approximation of the Kolmogorov complexity of the subarrays $r_u$ obtained by using the algorithmic Coding theorem~\cite{ctm}, $A(G)_{d\times d}$ represents the set with elements $(r_u,n_u)$, obtained when decomposing the adjacency matrix of $G$ into non-overlapping squares of size $d$ by $d$. In each $(r_u,n_u)$ pair, $r_u$ is one such square and $n_u$ its multiplicity (number of occurrences). From now on $BDM(G,d=4)$ will be denoted only by $K(G)$, but it should be taken as an approximation to $K(G)$ unless otherwise stated (e.g. when taking the theoretical true $K(G)$ value). Once the CTM is calculated, the BDM can be implemented as a lookup table and hence runs efficiently in linear time for non-overlapping fixed size submatrices.

\end{document}